\documentclass[12pt]{article}
\usepackage[margin=1in]{geometry}
\usepackage{natbib}
\usepackage{amsmath, amssymb}
\usepackage{graphicx}
\usepackage{authblk}
\usepackage{setspace}
\usepackage{multirow}
\usepackage{booktabs}
\usepackage{caption}
\usepackage{subcaption}
\usepackage{algpseudocode} 
\usepackage{algorithm2e} 
\begin{document}
\RestyleAlgo{ruled} 

\renewcommand\Authfont{\fontsize{12}{14.4}\selectfont}
\renewcommand\Affilfont{\fontsize{9}{10.8}\itshape}

\title{Distributed Lag Interaction Model with Index Modification}

\author[1]{Danielle Demateis}
\author[2]{Sandra India Aldana}
\author[2]{Robert O. Wright}
\author[3]{Rosalind Wright}
\author[4]{Andrea Baccarelli}
\author[2]{Elena Colicino}
\author[1]{Ander Wilson}
\author[1,*]{Kayleigh P. Keller}
\affil[1]{Department of Statistics, Colorado State University, Fort Collins, CO, USA}
\affil[2]{Department of Environmental Medicine and Climate Science, Icahn School of Medicine at Mount Sinai, New York, NY, USA }
\affil[3]{Department of Public Health, Icahn School of Medicine at Mount Sinai, New York, NY, USA}
\affil[4]{Harvard T.H. Chan School of Public Health, Boston, MA, USA}
\affil[*]{kayleigh.keller@colostate.edu}
\date{}
\maketitle

\begin{abstract}
Epidemiological evidence supports an association between exposure to air pollution during pregnancy and birth and child health outcomes. Typically, such associations are estimated by regressing an outcome on daily or weekly measures of exposure during pregnancy using a distributed lag model. However, these associations may be modified by multiple factors.  We propose a distributed lag interaction model with index modification that allows for effect modification of a functional predictor by a weighted average of multiple modifiers. Our model allows for simultaneous estimation of modifier index weights and the exposure-time-response function via a spline cross-basis in a Bayesian hierarchical framework. Through simulations, we showed that our model out-performs competing methods when there are multiple modifiers of unknown importance. We applied our proposed method to a Colorado birth cohort to estimate the association between birth weight and air pollution modified by a neighborhood-vulnerability index and to a Mexican birth cohort to estimate the association between birthing-parent cardio-metabolic endpoints and air pollution modified by a birthing-parent lifetime stress index.

\vspace{1em}

\end{abstract}

\maketitle

\doublespacing

\section{Introduction}
A wide body of literature supports an association between exposure to air pollution during pregnancy and birth and children's health outcomes \citep{sram_ambient_2005, bosetti_ambient_2010, warren_spatial-temporal_2012, stieb_ambient_2012, chang_assessment_2015, hsu_prenatal_2015, lakshmanan_associations_2015, chiu_prenatal_2016, jacobs_association_2017}. Studies often use a distributed lag model (DLM) to estimate the association between air pollution exposure and health outcomes \citep{chiu_prenatal_2016, hsu_prenatal_2015}. The DLM framework regresses the response onto repeated measures of exposure to form an exposure-time-response function of linear associations between the time-varying exposure and response \citep{almon_distributed_1965}. The exposure-time-response function can be used to identify windows of susceptibility, periods when there is an association. However, associations between environmental exposures and health outcomes can vary across populations \citep{arcaya_research_2016, martenies_associations_2022, schnake-mahl_gentrification_2020}. Therefore, effect modification in a DLM framework is of epidemiological interest.

Numerous extensions to DLMs exist to address methodological gaps by incorporating different forms of flexibility, including effect heterogeneity for pre-specified factors. \cite{wilson_bayesian_2017} developed a Bayesian distributed lag interaction model that includes interaction between the repeated measures of exposure and a single categorical variable. \cite{warren_spatial-temporal_2012} developed a spatially-varying Gaussian process model for windows of susceptibility that allows the exposure-time-response function to vary by areal units (e.g., census tracts). \cite{demateis_penalized_2024} developed a distributed lag interaction model (DLIM) to allow for heterogeneous effect modification across a population using a single continuous variable. A DLIM extends the exposure-time-response function to vary continuously across the population based on a single continuous modifier using a bi-dimensional function space or cross-basis \citep{demateis_penalized_2024, gasparrini_distributed_2010}. \cite{mork_heterogeneous_2024} developed a heterogeneous DLM that uses additive regression trees to estimate DLMs for subgroups of the population based on multiple continuous and categorical variables using a non-parametric regression tree method. In this paper, we consider extending the DLIM to multiple modifiers as a parsimonious and interpretable model.

Many studies have considered single or multiple index models to study the association between environmental mixture exposures and health outcomes. This is in part because linear index models are interpretable and can reduce the effects of multicolinearity that arise from correlated predictors.  The environmental health literature has widely adopted weighted quantile sum regression and quantile g-computation to evaluate the impact of environmental mixtures on health outcomes by creating a mixture index \citep{czarnota_assessment_2015, daniel_perinatal_2020, keil_quantile-based_2020, wheeler_assessment_2021, wu_occupational_2021} while more recent innovations have developed multiple index models \citep{mcgee_bayesian_2023}. Epidemiological researchers have also adopted indices for effect modification. For example, EnviroScreen metrics aggregate multiple continuous variables measured at the census-tract level into a single continuous index aiming to quantify environmental stressors in areas of residence. \cite{niu_association_2022} and \cite{martenies_associations_2022} used California EnviroScreen scores as modifiers of exposure-time-response functions. In other contexts, multiple measures of physical or behavioral health may be combined to form a risk index, e.g., a mental health index \citep{kopta_psychometric_2002} or a quality of life index \citep{kai_quality_1991}, to succinctly summarize multiple related factors. However, these analyses that consider modification through an index all rely on fixed, pre-specified index weights and do not allow for data-driven weight estimation or modifier selection within the index.  

In settings with repeated measures of exposure, creating a single index from many similar modifiers would allow for modification of the exposure-time-response function based on the cumulative impact of all modifiers. However, using an existing index with fixed weights can be problematic for two reasons. First, the fixed weights may not represent the true underlying structure of difference in susceptibility, leading to an incorrectly estimated exposure-time-response function. Second, this approach does not allow for identifying which of the candidate modifiers are driving the differences in the exposure-time-response function. Clearly, there is a need to simultaneously estimate modifier index weights and the exposure-time-response function in a distributed lag framework. 

We propose a distributed lag interaction model with index modification (DLIM-IM). The proposed approach extends the single modifier model of \cite{demateis_penalized_2024} to multiple modifiers through a data-derived modifier index. The methodology in this paper uses a Bayesian hierarchical framework to simultaneously estimate the exposure-time-response function and the index weightings for multiple continuous modifiers to create a single continuous index that modifies the exposure-time-response function. We applied our proposed method in two data analyses. The first uses a Colorado administrative birth cohort data set along with Colorado EnviroScreen to estimate the association between exposure to ambient $\text{PM}_{2.5}$ (particulate matter less than 2.5 microns in diameter) during gestation and birth weight for gestational age z-score (BWGAZ) modified by measures of neighborhood-level vulnerability. The second uses the Mexican Programming Research in Obesity, Growth, Environment, and Social Stressors (PROGRESS) cohort data set to estimate the association between exposure to ambient $\text{PM}_{2.5}$ surrounding pregnancy and birthing-parent cardio-metabolic endpoints modified by time after parturition. Our R package \texttt{dlimIM} is publicly available on GitHub. 

\section{Methods}

\subsection{Model}

For individual $i = 1, \dots, n$, let $y_i$ be the response, $\mathbf{x}_i = [x_{i1}, \dots, x_{iT}]'$ be a vector of exposures measured at time points $t = 1, \dots, T$, and $\mathbf{m}_i = [m_{i1}, \dots, m_{iL}]'$ be a vector of $L$ candidate modifying variables. We assume the modifying variables to be scaled to the unit interval and to operate in the same direction. Also, let $\mathbf{z}_i$ be a vector of $p$ covariates, including 1 for the intercept and the candidate modifier values so their main effects are included in the model. 

We make three key assumptions about the exposure-response relationship. First, we assume that the effect of exposure on the response at each time point is linear. Second, we assume that the exposure-response relationship varies smoothly across exposure-time. These first two assumptions are widely assumed across the DLM literature \citep{almon_distributed_1965}. Third, we assume that the relationship between repeated measures of the exposure and the response is modified by a linear combination of candidate modifiers.  

Let $\boldsymbol{\rho} = [\rho_1, \dots, \rho_L]'$ be the vector of index weights for the $L$ candidate modifiers such that $\rho_l \ge 0, l = 1,\dots,L$, and $\sum_{l=1}^L \rho_l = 1$. While other constraints have been used for single-index models, e.g., constraining one element of the index to be positive \citep{yu_penalized_2002} or $\text{L}_2$ constraint \citep{mcgee_bayesian_2023}, we chose to impose directional homogeneity because this is common for single-index models used in environmental epidemiology \citep{carrico_characterization_2015}. The weighted modifier index for individual $i$ is $m_i^* = \mathbf{m}_i'\boldsymbol{\rho}$. Our proposed model is
\begin{equation}\label{eq.main}
    g[E(y_i|\mathbf{x}_{i},\mathbf{z}_i, \mathbf{m}_i)]  = \sum_{t=1}^T x_{it}\beta_t(m_i^*) + \mathbf{z}_i'\boldsymbol{\gamma},
\end{equation}
where $g$ is the link function, $\beta_t(m_i^*)$ is the linear effect of exposure at time $t$ on the response for an individual with modifier index value $m_i^*$, and $\boldsymbol\gamma$ is a vector of regression coefficients for the covariates and intercept. When $m^*$ is replaced by a scalar modifying factor, the model in \eqref{eq.main} is the distributed lag interaction model presented by \cite{demateis_penalized_2024}. Here, we extend the model to include a data-driven index of multiple candidate modifiers.

From \eqref{eq.main}, we focus inference on three parameters. First, the exposure-time-response function for weighted modifier $m^*$ is $[\beta_1(m^*), \dots, \beta_T(m^*)]'$. The exposure-time-response function quantifies how the effect of exposure on response varies both as a function of exposure-time and the modifier index, which allows us to identify windows of susceptibility and ranges of the modifier index associated with susceptibility. We further discuss the parameterization of the exposure-time-response function in Section~\ref{sec.parameterize}. Second, the cumulative effect for an individual with modifier value $m^*$ is $\text{CE}(m^*) = \sum_{t=1}^T \beta_t(m^*)$. For a model with an identity link function, the cumulative effect is the expected aggregated change in response per unit of exposure at all time points across the exposure history. Third, the modifier index weights $\boldsymbol{\rho}$ provide insight into which modifiers contribute to altering the exposure-time-response function and their level of contribution. We propose methodology for a model with a Gaussian family and identity link function and a model with a binomial family and logit link function. 

\subsection{Parameterizing the Exposure-Time-Response Functions} \label{sec.parameterize}

We parameterize the exposure-time-response function as a smooth function of both exposure time and the weighted modifier index. We parameterize the exposure-time-response vector $[\beta_{1}(m_i^*), \dots, \beta_{T}(m_i^*)]$ using a cross-basis, which is a bi-dimensional function space that describes both the modified exposure-time-response associations and repeated measures across time. To create the cross-basis, we use splines to smooth in the modifier and exposure-time dimensions similar to the approach in \cite{demateis_penalized_2024}. We use natural splines that include an intercept to expand the weighted modifier index with $\nu_{mod}$ degrees of freedom. Scaling $[m_{1l}, \dots, m_{nl}]$ for each $l = 1, \dots, L$ to lie within the unit interval allows us to set boundary knots at 0 and 1 to ensure all possible modifier index values are within boundary knots. We denote the modifier index basis evaluated at value $m_i^*$ as $\mathbf{b}(m_i^*) = [b_1(m_i^*), \dots, b_{\nu_{mod}}(m_i^*)]'$ for $i=1,\dots,n$. We also use natural splines that include an intercept and boundary knots at the first and last exposure-time points to expand in the exposure-time dimension with $\nu_{time}$ degrees of freedom. Including an intercept in the modifier basis allows us to include a main effect of the exposure history. Without the intercept, a reference point for the modifier index is necessary for interpreting changes in the exposure-time-response function per modifier index value \citep{gasparrini_distributed_2010}. The basis evaluated at exposure-time $t$ for $t = 1, \dots, T$ is $\mathbf{c}(t) = [c_1(t), \dots, c_{\nu_{time}}(t)]'$. Combining these two basis expansions, the cross-basis for individual $i$ is $w_{jk}(m_i^*, \mathbf{x}_i) = \sum_{t=1}^T x_{it} b_k(m_i^*)c_j(t)$ for $k=1,\dots,\nu_{mod}$ and $j=1,\dots,\nu_{time}$. 

Using this parameterization, the sum $\sum_{t=1}^T x_{it}\beta_t(m_i^*)$ in \eqref{eq.main} becomes $\sum_{k = 1}^{\nu_{mod}} \sum_{j = 1}^{\nu_{time}} \theta_{jk} \times w_{jk}(m_i^*, \mathbf{x}_i)$. We denote the $n \times \nu_{time}\nu_{mod}$ design matrix for the cross-basis as $\mathbf{W}(\mathbf{X}, \mathbf{M}, \boldsymbol{\rho})$ and the $\nu_{time}\nu_{mod}$ cross-basis regression coefficients as $\boldsymbol{\theta}$. Let $\mathbf{U}(\boldsymbol{\rho}) = [\mathbf{W}(\mathbf{X}, \mathbf{M}, \boldsymbol{\rho}), \mathbf{Z}]$ and $\boldsymbol{\Psi} = [\boldsymbol{\theta}', \boldsymbol{\gamma}']'$. Under this parameterization, \eqref{eq.main} for all individuals is
\begin{equation}
    g[E\{\mathbf{y}|\mathbf{W}(\mathbf{X}, \mathbf{M}, \boldsymbol{\rho}),\mathbf{Z}, \mathbf{M}\}] = \mathbf{U}(\boldsymbol{\rho})\boldsymbol{\Psi}. 
\end{equation}

\subsection{Prior Specification}\label{sec.prior}

We consider prior specifications for modeling the modifier index weights $\boldsymbol{\rho}$ with and without selection. For the model without modifier selection, we specify a Dirichlet prior on the weights, $\boldsymbol{\rho} \sim \text{Dir}(\mathbf{q})$, where $\mathbf{q} = [q_1, \dots, q_L]'$ is a vector of hyper-parameters controlling the manner and strength of influence that the prior has on the posterior \citep{mcgee_incorporating_2023}. We parameterize the Dirichlet distribution using its relation to independent, normalized Gamma random variables. We let $a_l \sim \text{Gamma}(q_l, 1), l = 1, \dots, L$ and define $\rho_l = a_l / \sum_{l=1}^L a_l$. This ensures $\rho_l \ge 0$ for $l = 1, \dots, L$, $\sum_{l=1}^L \rho_l = 1$, and $\boldsymbol{\rho} \sim \text{Dir}(\mathbf{q})$. 

For a high-dimensional modifier space, we may be interested in a more parsimonious model via variable selection on the candidate modifiers. We again set $\rho_l = a_l / \sum_{l=1}^L a_l$, but now impose a spike-and-slab prior on the weights to perform selection on the modifiers included in the modification index \citep{koslovsky_bayesian_2023, mcgee_incorporating_2023}. The spike-and-slab prior 
\begin{equation} \label{eq.spikeslab}
    a_l | \eta_l \sim \eta_l \times \text{Gamma}(q_l,1) + (1 - \eta_l)\delta_0
\end{equation}
is a mixture distribution of a Dirac-delta function $\delta_0$ (spike) and a Gamma distribution (slab). The indicator for inclusion of weight $l$ in the prior is $\eta_l|\nu_l\sim$Bernoulli($\nu_l$), and $\nu_l$ is the prior inclusion probability. We fix $\nu_l = 0.5$. Adjusting $\nu_l$ controls sparsity, which may be helpful in low signal settings.

The Dirichlet prior for the weights can be made more or less informative through changes in $q_l$. A more informative Dirichlet prior could be centered on values based on prior information. For example, \cite{mcgee_incorporating_2023} center a Dirichlet prior for weights for components of an environmental mixture on toxicologically derived values, which could be helpful in low signal settings. A less informative Dirichlet prior puts equal weights on all modifiers and should be used if no prior information is available. 

We use a flat prior for the intercept, and we use independent normal priors with mean zero and variance $\tau^2$ for the cross-basis regression coefficients $\boldsymbol{\theta}$. For the covariate regression coefficients excluding the intercept, we use independent normal priors with mean zero and variance $\xi^2$. We assume an inverse-Gamma prior on the variance $\sigma^2$.

\subsection{Parameter Estimation}\label{sec.est}

We use Markov chain Monte Carlo (MCMC) to sample from the posterior distribution. The model has closed form full conditional distributions for regression coefficients $\boldsymbol{\Psi}$ and error variance but not for the modifier index weights. Therefore, we use a Metropolis-Hastings within Gibbs sampler to sample from the posterior distribution.  We sample regression coefficients $\boldsymbol{\Psi}$ and error variance $\sigma^2$ from a Gibbs sampler, and we sample the modifier index weights using a Metropolis-Hastings algorithm. The algorithm for our MCMC sampler is outlined in Supplemental Algorithm 1. For a logistic model, we use P\'olya-Gamma augmentation in a Gibbs sampler as in \cite{polson_bayesian_2013} for the regression coefficients and associated latent variables. See the Supplemental Section 1 for more details. 

For a model without modifier selection, we use the proposal distribution $a^{*}_l| a_l^{(s-1)} \sim \text{FoldedN}\left(a^{(s-1)}_l , \zeta_l^2 \right)$ to propose an update $a^{*}_l$ for the $l^{\text{th}}$ weight based on the current weight $a_l^{(s-1)}$ at iteration $s-1$, where $\zeta_l$ is an adaptively tuned hyperparameter. We accept the proposal $a_l^*$ with probability $\text{min}(r, 1)$, where 
\begin{equation} \label{eq.r}
    r = \frac{(a_l^*)^{q_l-1}\exp(-a^{*}_l)}{(a_l^{(s-1)})^{q_l-1}\exp(-a_l^{(s-1)})} \frac{p(\mathbf{y} | \mathbf{a}^*, \boldsymbol{\Psi}^{(s)}, \cdot)}{p(\mathbf{y} | \mathbf{a}^{(s-1)}, \boldsymbol{\Psi}^{(s)}, \cdot)}.
\end{equation}
Here, $\mathbf{a}^{(s-1)}$ is the current vector of un-normalized weights, $\mathbf{a}^*$ is the same as $\mathbf{a}^{(s-1)}$ with the $l^{\text{th}}$ weight replaced by $a^{*}_l$, and $\boldsymbol{\Psi}^{(s)}$ is the vector of regression coefficients at iteration $s$. In \eqref{eq.r}, $p$ is the normal density for a Gaussian model or the binomial density for a binomial model, where $\cdot$ represents other model parameters that are family-dependent (e.g., error variance for a Gaussian model). 

When performing modifier selection, we first propose $a^{*}_l$ as an update for un-normalized weight $l$ from 
\begin{equation*}
a^{*}_l| a_l^{(s-1)} \sim 
\left\{
    \begin{array}{ll}
        \text{Gamma}(q_l,1), & \text{if } a_l^{(s-1)} = 0 \\
        \delta_0, & \text{if } a_l^{(s-1)} \ne 0
    \end{array}
\right \}
\end{equation*}
and accept that proposed update with probability $\text{min}(r_{\text{select}}, 1)$, where 
\begin{equation}\label{eq.r_select}
r_{\text{select}} = 
\left\{
    \begin{array}{cl}
        \frac{1-\nu_l}{\nu_l} \frac{p(\mathbf{y} | \mathbf{a}^*, \boldsymbol{\Psi}^{(s)}, \cdot)}{p(\mathbf{y} | \mathbf{a}^{(s-1)}, \boldsymbol{\Psi}^{(s)}, \cdot)}, & \text{if } a_l^*=0, a_l^{(s-1)} \ne 0\\
        \frac{\nu_l}{1-\nu_l} \frac{p(\mathbf{y} | \mathbf{a}^*, \boldsymbol{\Psi}^{(s)}, \cdot)}{p(\mathbf{y} | \mathbf{a}^{(s-1)}, \boldsymbol{\Psi}^{(s)}, \cdot)}, & \text{if } a_l^*\ne 0, a_l^{(s-1)} = 0\\
    \end{array}
\right \}.
\end{equation}
If the proposal is $a^*=0$ and is rejected, we then propose a new non-zero value from $\text{FoldedN}\left(a^{(s-1)}_l, \zeta_l^2 \right)$ and accept or reject that value using the acceptance ratio in \eqref{eq.r}.

To obtain marginal posterior samples for the exposure-time-response associations and cumulative effects, we use linear transformations of the marginal posterior samples for $\boldsymbol{\theta}$ at each iteration $s$. To obtain posterior samples of the exposure-time-response associations for modifier index $m^*$, we transform the posterior samples of $\boldsymbol{\theta}$ at each iteration $s$ using the transformation $[\mathbf{b}(m^*) \otimes \mathbf{C}]\boldsymbol{\theta}^{(s)}$ to obtain $[\beta_1(m^*)^{(s)}, \dots, \beta_T(m^*)^{(s)}]'$. Here, $\mathbf{C}$ is the $T \times \nu_{time}$ exposure-time basis. To obtain posterior samples of the cumulative effect for modifier index $m^*$, we transform the posterior samples of $\boldsymbol{\theta}$ at each iteration $s$ using the transformation $\mathbf{w}_*(m^*) \boldsymbol{\theta}^{(s)}$ to obtain $\text{CE}(m^*)^{(s)}$. Here, $\mathbf{w}_*(m^*)$ is the vectorization, by column, of $[\mathbf{1}' \otimes \mathbf{b}(m^*)]\mathbf{C}$. We use the posterior means and intervals of these transformed parameters to perform inference.  

\section{Simulation Study} 

We evaluated the proposed DLIM-IM's ability to estimate cumulative effects and exposure-time-response functions, as well as identify correct modifier index weights. We provide results for our proposed DLIM-IM with and without modifier selection using 5 degrees of freedom in both the modifier and exposure-time dimensions. For prior distributions, we use $q_l = 1$ for all $l$ for the weights prior, $\tau^2 = 100$ for the cross-basis regression coefficients' priors, $\xi^2 = 110$ for the covariate regression coefficients' priors, and we use 1 and 0.001 for the shape and scale parameters of the error variance prior. We compare to results from a penalized DLIM with 20 basis functions in each dimension as described in \cite{demateis_penalized_2024}, fixing the weights equally in each scenario, which we refer to as the fixed-index model. Code for implementing this simulation is provided as a companion file to this manuscript and is available on GitHub.

\subsection{Simulation Design and Data Generation}\label{sec.sim}

We simulated data using $n=1000$ vectors of real $\text{PM}_{2.5}$ concentration measurements over 37 weeks. We generated modifier values from multivariate normal distributions with zero means, variances of 1, and covariances ranging from 0.4 to 0.7. To do this, we generated a covariance matrix with 1s on the diagonal and sampled from a uniform distribution on the off-diagonal elements, forcing the matrix to be both symmetric and positive definite. For each simulated data set, we scaled each of the $L$ modifiers for individuals $i = 1, \dots, n$ to range from 0 to 1. 

We considered three modifier index weighting scenarios. Scenario 1 uses equal weights $\rho_1 = \rho_2 = \rho_3 = 1/3$ as a baseline for comparison to a fixed-index. Scenario 2 has 3 different weights (0.5, 0.4, 0.1). Scenario 3 has 50 weights, 3 of which are 1/3 and the rest are zero. See Supplemental Table 1 for more weight scenarios. 

We generated 3 covariates from multivariate normal distributions with the mean of the first week of exposure as the mean and the following covariance structure: $\text{cov}(\mathbf{z}_1, \mathbf{z}_2) = 0.5$, $\text{cov}(\mathbf{z}_1, \mathbf{z}_3) = 0.6$, $\text{cov}(\mathbf{z}_2, \mathbf{z}_3) = 0.7$, $\text{Var}(\mathbf{z}_1) = \text{Var}(\mathbf{z}_2) = \text{Var}(\mathbf{z}_3) = 1$. We generated regression coefficients for the modifiers from a uniform distribution on $(-1,1)$ and regression coefficients for the covariates from independent standard normal distributions. The exposure-time-response function for the data generating mechanism was $\beta_t(m_i^*)=m_i^* \times f(t,37[1+\exp\{-20(m_i^*-0.5)\}]^{-1})$, where $f(t,c) := 2.5\phi[(t-c)/5]$ and $\phi(\cdot)$ is the normal probability density function. We simulated response values using a normal distribution with variance $\sigma^2$. We considered three signal-to-noise settings: low (0.1), medium (0.5), and high (1). The signal is the standard deviation of the simulated exposure-time-response function, and noise is $\sigma$. We simulated 200 data sets for each combination of weight scenarios and signal-to-noise settings. See Supplemental Figure 1 for the scaling of computation time as a function of sample size.

\subsection{Performance Evaluation}\label{sec.performance}

To evaluate modifier index weight estimation, we provide box plots for weight estimates across 200 simulated data sets for our proposed models. In Supplemental Figure 2, we provide posterior inclusion probabilities (PIPs) for the modifier index weights from our proposed modifier selection models. Using the estimated modifier index weights, we constructed the weighted modifier index for each simulated data set and computed the root mean squared error (RMSE) as $[n^{-1}\sum_{i=1}^n \{m^*_i - \hat{m}^*_i\}^2]^{1/2}$ and average absolute bias as $n^{-1}\sum_{i=1}^n |m^*_i - \hat{m}^*_i|$, where $\hat{m}^*_i = \mathbf{m}'_i \hat{\boldsymbol{\rho}}$ and $\hat{\boldsymbol{\rho}}$ is the posterior mean of $\boldsymbol{\rho}$. We averaged both RMSE and average absolute bias over the 200 simulated data sets.

We also evaluated average RMSE, coverage, and credible or confidence interval (CI) width for effect estimates. For cumulative effect estimates for each data set, we calculated RMSE as  $[n^{-1}\sum_{i=1}^n \{\text{CE}(m_i) - \hat{\text{CE}}(m_i)\}^2]^{1/2}$, and we calculated coverage as  $n^{-1}\sum_{i=1}^n \mathbb{I}[Q_{\alpha/2}\{\text{CE}(m_i) | \mathbf{y}\} \allowbreak \le \text{CE}(m_i) \le Q_{1-\alpha/2}\{\text{CE}(m_i) | \mathbf{y}\}]$ for credible intervals and as $n^{-1}\sum_{i=1}^n \mathbb{I}[|\text{CE}(m_i) - \hat{\text{CE}}(m_i)| \allowbreak \le \Phi^{-1}(1-\alpha/2) \hat{\text{var}} \{ \hat{\text{CE}}(m_i) \}]$ for confidence intervals. The function $\mathbb{I}(A)$ is an indicator of event $A$'s occurrence, and $Q_{a}(.|\mathbf{y})$ is the $a^{th}$-quantile of the sampled posterior distribution for a given parameter. For point-wise effect estimates for each data set, we calculated RMSE as $[n^{-1} T^{-1} \sum_{i=1}^n    \sum_{t=1}^T \{\beta_t(m_i) - \hat\beta_t(m_i)\}^2]^{1/2}$, and we calcualted coverage as $n^{-1} T^{-1} \sum_{i=1}^n  \sum_{t=1}^T \mathbb{I}[Q_{\alpha/2}\{\beta_t(m_i) | \mathbf{y}\} \le \beta_t(m_i) \le Q_{1-\alpha/2}\{\beta_t(m_i) | \mathbf{y}\}]$ for credible intervals and as $n^{-1} T^{-1} \sum_{i=1}^n  \sum_{t=1}^T \mathbb{I}[|\beta_t(m_i) - \hat{\beta}_t(m_i)| \le \Phi^{-1}(1-\alpha/2) \hat{\text{var}} \{ \hat{\beta}_t(m_i) \}]$ for confidence intervals. We also compared average credible interval width and average confidence interval width for the proposed and fixed-index models, respectively. For cumulative effect estimates, we calculated average credible interval width as $n^{-1}\sum_{i=1}^n [Q_{1-\alpha/2}\{\text{CE}(m_i) | \mathbf{y}\} - Q_{\alpha/2}\{\text{CE}(m_i) | \mathbf{y}\}]$ and average confidence interval width as $n^{-1}\sum_{i=1}^n  2\Phi^{-1}(1-\alpha/2) [\hat{\text{var}} \{ \hat{\text{CE}}(m_i) \}]^{1/2}$  for each data set. For point-wise effect estimates, we calculated average credible interval width as $n^{-1} T^{-1} \sum_{i=1}^n  \sum_{t=1}^T [Q_{1-\alpha/2}\{\beta_t(m_i) | \mathbf{y}\} - Q_{\alpha/2}\{\beta_t(m_i) | \mathbf{y}\}]$ and average confidence interval width as $n^{-1} T^{-1} \sum_{i=1}^n  \sum_{t=1}^T 2\Phi^{-1}(1-\alpha/2) [\hat{\text{var}} \{ \hat{\beta}_t(m_i) \}]^{1/2}$. We averaged RMSE, coverage, and average CI width across the 200 simulated data sets.

\subsection{Simulation Results} \label{sec.simresults}

\begin{figure*}[!t]%
    \centering
    \includegraphics[width=1\textwidth]{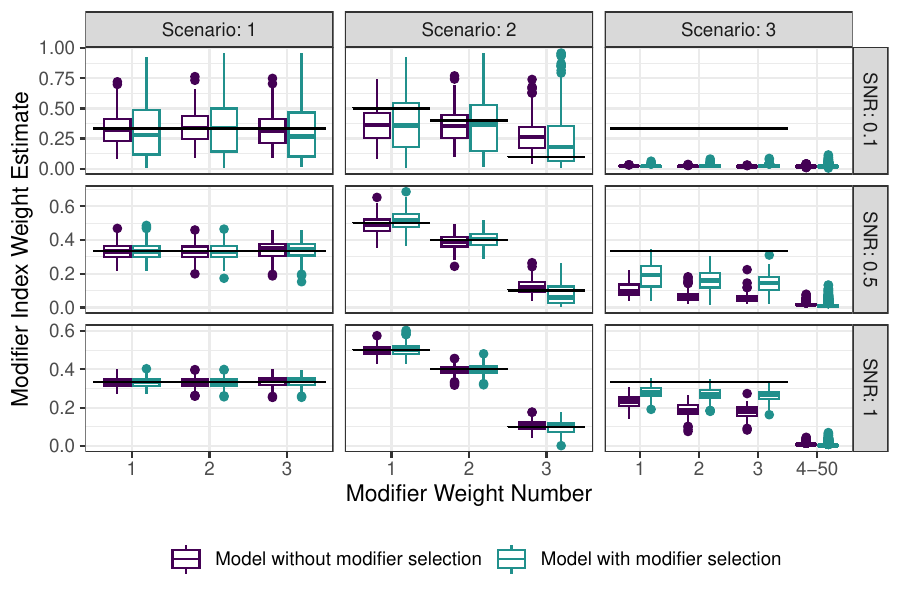}
    \caption{Modifier index weight estimates from DLIM-IMs with selection and DLIM-IMs without selection for all 200 simulated data sets across all weight scenarios and signal-to-noise (SNR) settings. The number of basis functions for the exposure-time basis is 5 for both the DLIM-IMs. The black line represents the true weight value. For scenario 3 (sparse weight scenario), weights 1-3 are non-zero, and weight numbers 4-50 is averaged over all 47 zero weights.}
    \label{fig:weights}
\end{figure*}

Figure~\ref{fig:weights} shows summaries of the modifier index weight estimates. Table~\ref{tab:main} includes a summary of the cumulative and point-wise effect estimates for the scenarios in our simulation study. In the Supplemental Figures 3-5, we provide figures of the estimated exposure-time-response curve for individual data sets. Our proposed DLIM-IM produces more accurate modifier index weight estimates than assuming equal weights, and the proposed DLIM-IM with modifier selection is able to identify factors that modify the distributed lag function. The DLIM-IMs performed better than the fixed-index model in terms of cumulative effect performance metrics and performed better in terms of some point-wise performance metrics as well. 

\begin{table*}[!t] 
\caption{Performance metrics for the weighted modifier index, cumulative effects, and point-wise effect for DLIM-IMs without selection (DLIM-IM), the DLIM-IM with selection (DLIM-IM sel.), and the fixed-index model (fixed-index) averaged over all 200 simulated data sets with non-linear modification for all three weight scenarios and three signal-to-noise (SNR) levels. The low SNR is 0.1, medium SNR is 0.5, and the high SNR is 1. The number of basis functions for the exposure-time basis was 5 for both DLIM-IMs and 20 for the fixed-index model.}
\tabcolsep=0pt
\label{tab:main}
\begin{tabular*}{\textwidth}{@{\extracolsep{\fill}}ccccccccc@{\extracolsep{\fill}}}
\toprule%
& & Index & \multicolumn{3}{@{}c@{}}{Cumulative Effect} & \multicolumn{3}{@{}c@{}}{Point-wise Effect} \\
\cline{3-3}\cline{4-6}\cline{7-9}%
SNR & Model & RMSE &  RMSE  & Coverage & Width & RMSE & Coverage & Width \\ 
\midrule

        \multicolumn{9}{@{}c@{}}{Scenario 1: Equal Weights}\\
        \midrule
        
         & DLIM-IM & 0.0221 & 1.10 & 0.95 & 4.96 & 0.23 & 0.96 & 0.88 \\ 
        low & DLIM-IM sel. & 0.0381 & 1.09 & 0.95 & 4.96 & 0.23 & 0.96 & 0.88 \\
         & Fixed-index & 0 & 1.77 & 0.94 & 7.29 & 0.12 & 0.92 & 0.42 \\   \hline
         & DLIM-IM & 0.0081 & 0.28 & 0.96 & 1.29 & 0.06 & 0.89 & 0.19 \\ 
        med. & DLIM-IM sel. & 0.0082 & 0.28 & 0.96 & 1.29 & 0.06 & 0.89 & 0.19 \\ 
         & Fixed-index & 0 & 0.36 & 0.95 & 1.51 & 0.04 & 0.72 & 0.10 \\   \hline
         & DLIM-IM & 0.0041 & 0.18 & 0.95 & 0.74 & 0.04 & 0.77 & 0.10 \\ 
        high & DLIM-IM sel. & 0.0041 & 0.18 & 0.95 & 0.74 & 0.04 & 0.77 & 0.10 \\ 
         & Fixed-index & 0 & 0.19 & 0.95 & 0.79 & 0.03 & 0.87 & 0.09 \\  \hline
        
       \multicolumn{9}{@{}c@{}}{Scenario 2: Different Weights}\\
        \midrule
        
         & DLIM-IM & 0.0295 & 1.11 & 0.95 & 4.98 & 0.23 & 0.96 & 0.89 \\ 
        low & DLIM-IM sel. & 0.0398 & 1.10 & 0.95 & 4.98 & 0.23 & 0.96 & 0.89 \\
         & Fixed-index & 0.0273 & 1.78 & 0.94 & 7.32 & 0.12 & 0.92 & 0.42 \\  \hline
         & DLIM-IM & 0.0078 & 0.28 & 0.96 & 1.27 & 0.06 & 0.89 & 0.19 \\ 
        med. & DLIM-IM sel. & 0.0094 & 0.27 & 0.96 & 1.26 & 0.06 & 0.89 & 0.18 \\ 
         & Fixed-index & 0.0273 & 0.37 & 0.95 & 1.53 & 0.04 & 0.70 & 0.10 \\   \hline
         & DLIM-IM & 0.0042 & 0.18 & 0.95 & 0.73 & 0.04 & 0.77 & 0.10 \\  
        high & DLIM-IM sel. & 0.0049 & 0.18 & 0.95 & 0.73 & 0.04 & 0.77 & 0.10 \\
         & Fixed-index & 0.0273 & 0.20 & 0.95 & 0.83 & 0.03 & 0.77 & 0.08 \\ \hline
        
        \multicolumn{9}{@{}c@{}}{Scenario 3: Sparse Weights}\\
        \midrule 
        
         & DLIM-IM & 0.0603 & 1.18 & 0.98 & 6.19 & 0.26 & 0.96 & 0.97 \\ 
        low & DLIM-IM sel. & 0.0600 & 1.12 & 0.98 & 6.04 & 0.26 & 0.96 & 0.99 \\ 
         & Fixed-index & 0.0606 & 1.78 & 0.96 & 7.76 & 0.13 & 0.91 & 0.45 \\  \hline
         & DLIM-IM & 0.0485 & 0.36 & 0.96 & 1.63 & 0.07 & 0.89 & 0.22 \\ 
        med. & DLIM-IM sel. & 0.0333 & 0.32 & 0.97 & 1.52 & 0.07 & 0.90 & 0.21 \\
         & Fixed-index & 0.0606 & 0.39 & 0.94 & 1.66 & 0.05 & 0.67 & 0.10 \\   \hline
         & DLIM-IM & 0.0255 & 0.20 & 0.94 & 0.84 & 0.04 & 0.80 & 0.11 \\ 
        high & DLIM-IM sel. & 0.0129 & 0.18 & 0.94 & 0.79 & 0.04 & 0.79 & 0.10 \\ 
         & Fixed-index & 0.0606 & 0.23 & 0.93 & 0.95 & 0.04 & 0.59 & 0.07 \\ 
        
    \bottomrule
\end{tabular*}
\end{table*}

The proposed DLIM-IM produces more accurate weight estimates than assuming equal weights as shown in Figure~\ref{fig:weights}, which translates to more accurate estimates of the weighted modifier index. In scenarios 2 and 3 where the fixed model weights are mis-specified, the RMSE and absolute bias for the estimated weighted modifier index are better for the DLIM-IMs compared to the fixed-index model. In the high signal-to-noise setting for scenario 2, the RMSE was 0.004 for the DLIM-IMs and 0.027 for the fixed index model, and the absolute bias was 0.003 for the DLIM-IMs and 0.022 for the fixed-index model. In scenario 3, the DLIM-IM with selection produced more accurate weighted  modifier index estimates than the DLIM-IM without selection. In the high signal-to-noise setting, the RMSE was 0.0129 for the selection model and 0.0255 for the model without selection. The absolute bias was 0.0103 for the selection model and 0.0204 for the model without selection. The model with selection produces PIPs above 0.5 for non-zero weights and below 0.5 for zero weights. Our results demonstrate that the proposed DLIM-IM can estimate modifier index weights and select modifiers in various settings. 

The DLIM-IMs performed better than the fixed-index model in terms of cumulative effect estimate metrics. In all three scenarios, cumulative coverage was similar among the three models, but the DLIM-IMs had smaller CI widths on average. For example in the high signal-to-noise setting for scenario 1, the cumulative coverage was 0.95 for the three models; however, the average CI width was 0.74 for the fixed-index model and 0.79 for both DLIM-IMs with and without selection. In all three scenarios, cumulative RMSE was similar or lower for the DLIM-IMs than the fixed-index model. In scenario 3, the DLIM-IM with selection had better performance metrics than the DLIM-IM without selection. For example in the medium signal-to-noise setting for scenario 3, the average CI width for the DLIM-IM with and without selection was 1.52 and 1.63, respectively, and the RMSE was 0.32 and 0.36, respectively. Our results show that the DLIM-IM can estimate cumulative effects more efficiently and with coverage closer to the nominal level than the fixed-index model, and the DLIM-IM with selection estimates cumulative effects better than the DLIM-IM without selection when weights are sparse. 

The DLIM-IMs performed better than the fixed-index model in terms of point-wise effect estimate metrics. In all three scenarios and signal-to-noise settings, except the high signal-to-noise setting for scenario 1, the DLIM-IM had point-wise coverage similar or closer to the nominal level than the fixed-index model. For example in the medium signal-to-noise setting for scenario 2, the point-wise coverage was 0.89 for the DLIM-IMs with and without selection and 0.70 for the fixed-index model. However, point-wise RMSE was either similar or lower for the fixed-index model. We attribute this to the penalization of splines in the fixed-index model. In scenario 3, point-wise coverage, RMSE, and average CI width were similar for the DLIM-IMs. Overall, the DLIM-IMs had point-wise effect coverage closer to the nominal level than the fixed-index model and RMSE and average CI width similar to or better than the fixed-index model, while performance metrics for the two DLIM-IMs were similar.

\section{Data Analyses} \label{sec. analyese}

We used our proposed model to analyze two data sets. In the first application, we demonstrate the DLIM-IM's ability to perform selection on multiple continuous modifiers instead of including all modifiers with fixed weights. We use our proposed model with selection to extend the analysis of Colorado administrative birth data performed in \cite{demateis_penalized_2024}, which used a pre-defined index of 15 modifiers. Here, we extend the analysis to identify which factors contribute to modification of the association between exposure to ambient $\text{PM}_{2.5}$ during pregnancy and BWGAZ. In the second application, we demonstrate the DLIM-IM's ability to combine multiple measures of birthing-parent stress during gestation and lifetime instead of performing separate analyses on each modifier or arbitrarily choosing weights. We use the DLIM-IM without selection to create a weighted lifetime stress index using 4 birthing-parent stress variables. We use this lifetime stress index as a modifier of the association between exposure to ambient $\text{PM}_{2.5}$ surrounding pregnancy and cholesterol levels from the PROGRESS cohort data set. See Supplemental Figures 6 and 7 for information on modifier correlation for each analysis. For prior distributions, we use $\tau^2 = 100$ for the cross-basis regression coefficients priors, $\xi^2 = 110$ for the covariate regression coefficients priors, and we use 1 and 0.001 for the shape and scale parameters of the error variance prior.

\subsection{Air Pollution, EnviroScreen and Birth Weight in Colorado}\label{sec.COES}

We used a Colorado administrative birth cohort data set containing all live, full-term, singleton births with dates of conception from 2007-2018 in census tracts in the Colorado Front Range with elevation lower than 6,000 feet \citep{demateis_penalized_2024}. We matched the data with weekly average ambient $\text{PM}_{2.5}$ concentration during the gestational period obtained from the down-scaled models published by the US Environmental Protection Agency (www.epa.gov/hesc/rsig-related-downloadable-data-files) at birthing-parent residences. We regressed BWGAZ, constructed using Fenton growth charts \citep{fenton_new_2003}, onto repeated measures of ambient $\text{PM}_{2.5}$ measured for the first 37 weeks of gestation. The final data set we used included 393,204 births in 786 census tracts across 13 counties.

For candidate modifiers, we used the continuous indicators included in the Colorado EnviroScreen health and social factors (HSF) score, excluding the indicator for low birth weight \citep{colorado_department_of_public_health_and_environment_cdphe_colorado_2022}. The HSF score from Colorado EnviroScreen is constructed using the following tract-level percentile variables: percent of people of color, disability rate, percent of households linguistically isolated, percent of households that are housing-cost burdened, percent of people with less than a high school education, percent of people with low income, diabetes prevalence, asthma hospitalization rate, rate of poor mental health, percent of people over 64 years of age, percent of people under 5 years of age, cancer prevalence, rate of heart disease in adults, and estimated life expectancy.  Indicators are directionally adjusted following standard practice within Colorado Enviroscreen. For example, the percentile values for average life expectancy were reversed so that a higher percentile corresponds with shorter life expectancy. We divide all candidate modifiers by 100 to the unit interval. We also include all candidate modifiers as covariates in the model by including the linear main effect of each component of EnviroScreen. In addition, we control for each pregnant person's age, body mass index (BMI), height, and weight at conception. We also control for each pregnant person's race and ethnicity, education, income, and marital status before pregnancy and self-reported prenatal care habits and smoking habits during pregnancy.

The analysis in \cite{demateis_penalized_2024} fit a DLIM using the HSF score constructed by geometrically averaging the candidate modifiers \textit{a priori} as a single continuous modifier and presented estimated exposure-time-response functions for the HSF score quartiles: 32, 46, and 59. We fit our proposed model and estimated the weights of the candidate modifiers constructing the HSF score to 1) compare estimated exposure-time-response functions using fixed and estimated weights for the candidate modifiers and 2) identify which of the 14 candidate modifiers contributed to modification of the exposure-time-response function. 

We fit our proposed model with modifier selection using 3 degrees of freedom in both the exposure-time and modifier dimensions (Supplemental Figures 8 and 9 and Supplemental Table 2), and we used $q_1 = \dots = q_{14} = 1$ as the hyperparameter for the Dirichlet prior. We ran our MCMC sampler with 5 chains with a 2.4GHz AMD Milan processor with 3.8GB of RAM for 50,000 iterations and discarded the first 30,000 for burn-in based on visual inspection of the parameters' trace plots. The runtime ranged from 3-5 days for each chain. We compare to the penalized DLIM with 20 B-spline basis functions in the exposure-time dimension and a linear modifier basis fit in \cite{demateis_penalized_2024}. We refer to this as the fixed-index model. We compare exposure-time-response functions estimated at Colorado EnviroScreen HSF scores to  corresponding estimated modifier index value. For example, we compare the exposure-time-response curve estimated for Colorado EnviroScreen HSF score 32 to the estimated curve at estimated modifier index score of 0.32. We refer to both the Colorado EnviroScreen HSF score 32 and estimated modifier index value 0.32 as HSF 32 for simplicity. 

\begin{figure*}[!t]
    \centering
    \subfloat[Estimated cumulative effects across health and social factor modifier values.]
    { \includegraphics[width=0.75\textwidth]{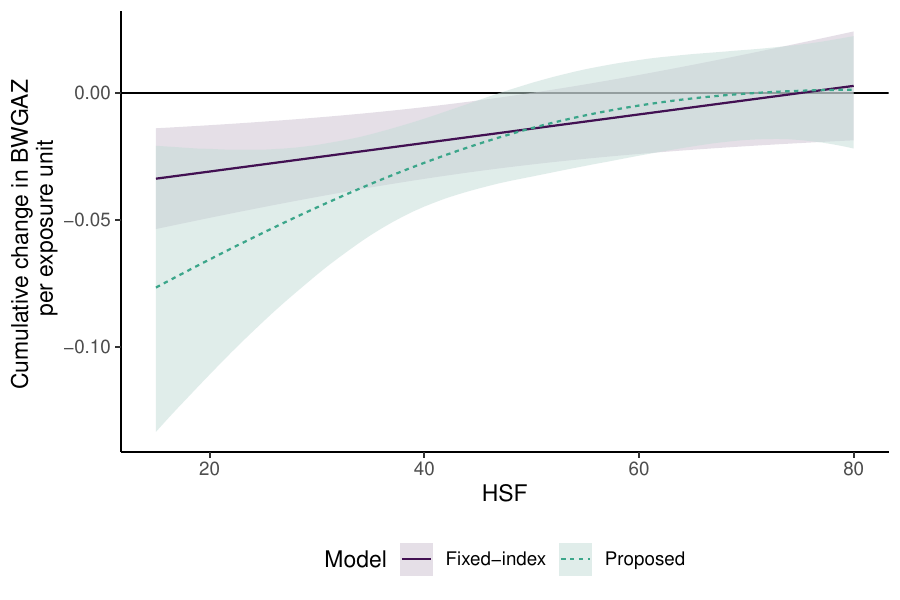} 
    \label{subfig:ES_cumul} }
    \hfill
    \subfloat[Estimated exposure-time-response curves across 37 weeks of gestation for the quartiles of the Colorado EnviroScreen health and social factors scores.]
    { \includegraphics[width=0.75\textwidth]{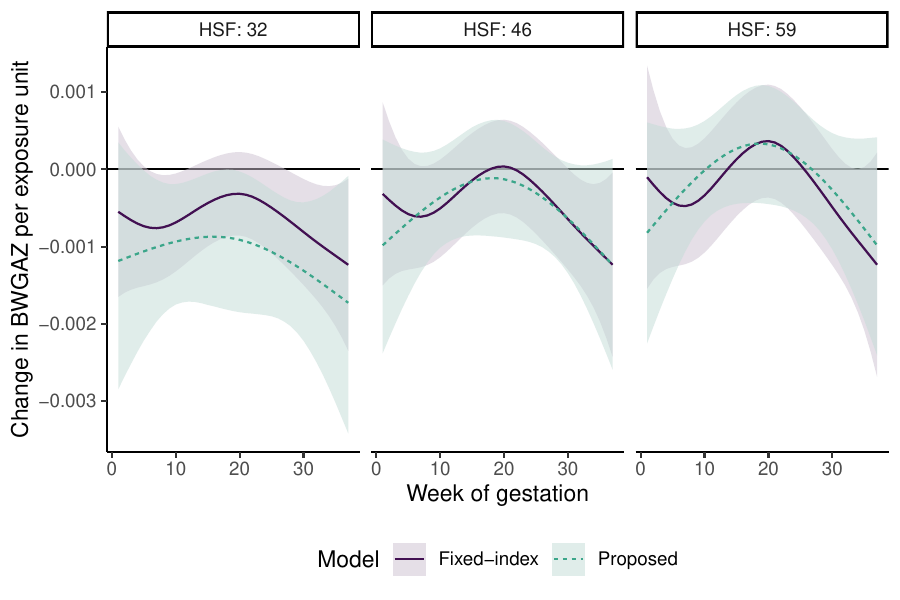}
    \label{subfig:ES_pw} }
    \caption{Estimated cumulative effect and exposure-time-response curves from the fixed-index penalized DLIM with linear modification and proposed DLIM-IM with selection applied to the Colorado birth cohort. The fixed-index DLIM uses 20 basis functions in the modifier and exposure-time dimensions, and the DLIM-IM with selection uses 5 basis functions in each dimension.}
    \label{fig:ES}
\end{figure*}

We estimated similar cumulative effects and different windows of susceptibility using both models. Figure~\ref{fig:ES} shows the cumulative effect estimates and estimated exposure-time-response curves from both the fixed-index and the proposed models. We estimated a negative cumulative effect of exposure for individuals with HSF scores below 46 using the proposed model and below 50 using the fixed-index model. Using the proposed model with data-derived modifier index, we estimated a slightly larger negative effect, although the intervals are wider and overlapping. We also estimated similar windows of susceptibility using both models. For HSF 32 with the fixed-index model, we estimated a window during the first and third trimester; whereas with the proposed model, we estimated a window from weeks 5 to 18 and another window from weeks 20-37. For HSF 46, we estimated a window during the third trimester with the fixed-index model, and we did not estimate any windows with the proposed model due to larger intervals in the more complex index modification model. For HSF 59, we did not estimate any windows with either model. 

Using the DLIM-IM, we were able to additionally identify which of the candidate modifiers contribute to modification. Table~\ref{tab:analysis} shows that 5 of the 14 candidate modifiers were selected into the model with PIPs greater than 0.5. Percent of people of color ($\hat{\rho} = 0.220$, $\text{PIP} = 0.807$), percent of people with less than high school education ($\hat{\rho} = 0.194$, $\text{PIP} = 0.749$), and asthma hospitalization rate ($\hat{\rho} = 0.135$, $\text{PIP} = 0.743$) were the top three factors contributing to modification. This indicates that these variables contributed most, per unit change, to the modification of the effect of exposure to $\text{PM}_{2.5}$ during pregnancy on BWGAZ. Because the modifier index weights are different from one another, this means that each of the modifiers has a different contribution to modification of the exposure-time-response function. While modifiers that were selected into the model (PIP $>0.5$) generally had estimated weights larger than the average of equal weights ($1/14$), and modifiers not selected generally had weight estimates that were smaller, we still use PIPs as a basis of selection because it is plausible that this is not always the case. See Supplemental Figures 10-12 for a comparison of the estimated exposure-time response curve from the model presented here and models with only one of the modifiers. Our results demonstrate how the proposed model can be used in an analysis with multiple modifiers to identify important factors that contribute to modification of the exposure-time-response function.

\begin{table*}[!t]
\centering 
\caption{Modifier index weight estimates, standard deviations, and posterior inclusion probabilities (PIP) for each of the 14 included indicators of the EnviroScreen HSF score obtained in the Colorado birth cohort analysis using our proposed model with 3 degrees of freedom for both the modifier and exposure-time bases. Results are summarized across 5 chains. Indicators marked with (*) have PIPs over 0.5, suggesting that these indicators contribute to modification of the exposure-time-response relationship. } 
\label{tab:analysis}
\begin{tabular*} {\textwidth}{@{\extracolsep{\fill}}cccc@{\extracolsep{\fill}}}
\toprule%

        Indicator & Posterior mean & Posterior SD & PIP \\ \midrule

        \% of people of color (*) & 0.220 & 0.179 & 0.807 \\ 
        \% of people w/ less than high school education (*) & 0.194 & 0.192 & 0.749 \\ 
        Asthma hospitalization rate (*) & 0.135 & 0.119 & 0.743 \\ 
        \% of people over age 64 (*) & 0.097 & 0.131 & 0.591 \\ 
        \% of households linguistically isolated (*) & 0.069 & 0.100 & 0.531 \\ 
        \% of households housing-cost burdened & 0.060 & 0.095 & 0.495 \\ 
        Disability rate & 0.055 & 0.097 & 0.460 \\  
        Diabetes prevalence & 0.046 & 0.071 & 0.472 \\
        Rate of poor mental health  & 0.029 & 0.059 & 0.374 \\
        Cancer prevalence & 0.034 & 0.059 & 0.410 \\ 
        \% of people under age 5 & 0.020 & 0.047 & 0.308 \\ 
        Rate of heart disease in adults & 0.015 & 0.047 & 0.236 \\ 
        Life expectancy & 0.013 & 0.034 & 0.254 \\ 
        \% of people with low income & 0.013 & 0.034 & 0.261 \\
        
        \bottomrule
\end{tabular*}
\end{table*}

\subsection{Air Pollution, Maternal Stress, and Lipids in PROGRESS}\label{sec.stress}

The PROGRESS cohort data set is a longitudinal birth cohort study with a total of 948 birthing-parent-infant pairs in Mexico City, Mexico that were recruited between 2007 and 2011 \citep{braun_relationships_2014, burris_association_2014}. We obtained daily $\text{PM}_{2.5}$ concentrations based on a validated model as described in \cite{gutierrez-avila_prediction_2022} at each of the participant's residential addresses and aggregated the concentrations monthly.  As responses, we used high-density lipoprotein (HDL) measured in plasma by enzymatic photometric assays and low-density lipoprotein (LDL) derived from the Friedewald equation \citep{friedewald_estimation_1972}. Using a DLIM-IM, we regressed birthing-parent HDL and LDL,  measured 4 years after parturition onto monthly average $\text{PM}_{2.5}$ starting at 2 months before through 22 months after each birthing-parent's last menstrual period (LMP). We controlled for each birthing-parent's age, BMI, assessed socio-economic status (lower, medium, and higher), and marital status at the second trimester. We also controlled for their smoking habits (passive or active) and parity at baseline, alcohol intake and cardiometabolic medications 2 years after parturition, and the season (November to February, March and April, and May to October) during LMP. After removing observations with incomplete data, the number of observations is 272.

We used 4 different measures of birthing-parent stress to create a data-derived lifetime stress index as a modifier of the exposure-time-response function. The stress modifiers we used are the Edinburgh Postnatal Depression scale (EPDS) as a measure of depression during gestation \citep{levis_accuracy_2020}, the Crisis in Family Systems (CRISYS) to quantify the impact of negative life events \citep{sherlock_short_2023}, the Perceived Stress Scale as a measure of self-reported stress during gestation \citep{cohen_global_1983}, and the State-Trait Anxiety Inventory (STAI) as a measure of anxiety during gestation \citep{spielberger_manual_1983}. Higher values for each modifier correspond to higher stress levels. Each of the four modifiers was measured once for each individual during gestation. To create a common scale across stress variables we rescale each modifier to range from 0 to 1.

We fit a DLIM-IM without selection using 3 degrees of freedom in both the exposure-time and modifier dimensions, and we used $q_1 = \dots = q_4 = 5$ as the hyperparameter for the Dirichlet prior. We ran our MCMC sampler with 5 chains for 30,000 iterations and discarded the first 10,000 for burn-in based on visual inspection of the trace plots, which had a runtime of about 5 minutes for each chain. We present results in the main text for only the HDL response as we did not estimate any effect of exposure to $\text{PM}_{2.5}$ on LDL but present results in Supplemental Figures 13 and 14. The weighted lifetime stress index we estimate combines various indicators of birthing-parent stress, with higher values indicating more stress. 

\begin{figure*}[!t]%
    \centering
    \includegraphics[width=1\textwidth]{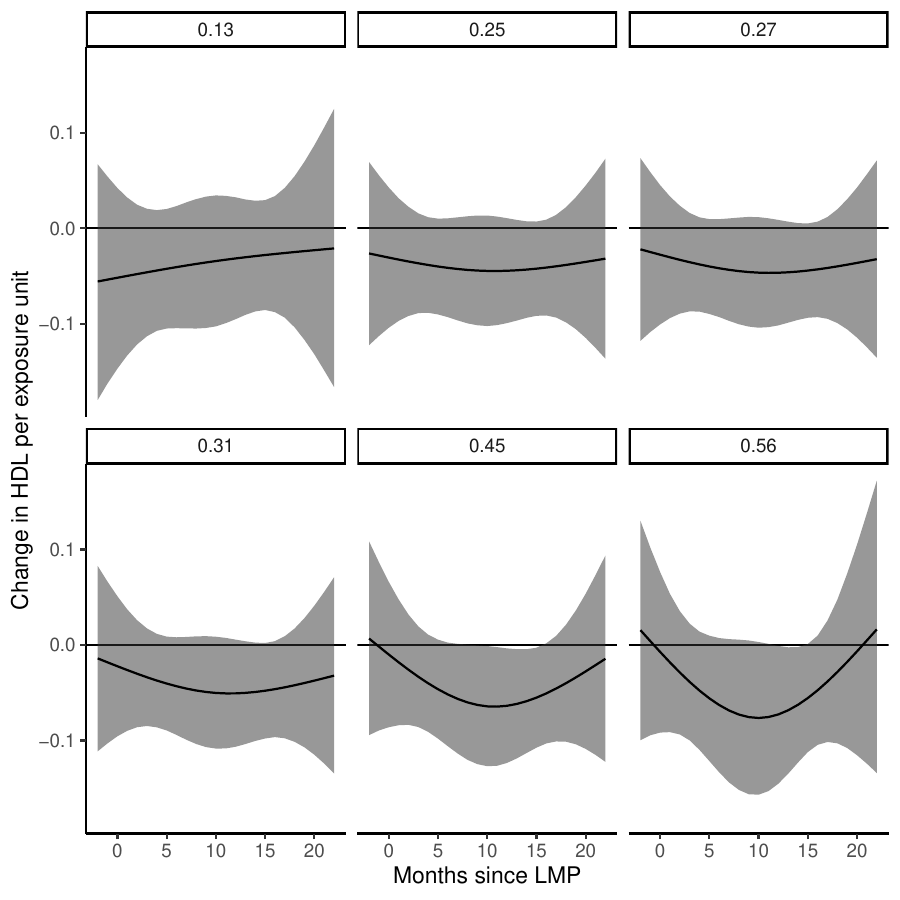}
    \caption{Estimated exposure-time-response curves for the $10^{\text{th}}$ ($\hat{m}^* = 0.13$), $25^{\text{th}}$ ($\hat{m}^* = 0.25$), $30^{\text{th}}$ ($\hat{m}^* = 0.27$), $40^{\text{th}}$ ($\hat{m}^* = 0.31$), $75^{\text{th}}$ ($\hat{m}^* = 0.45$), and $90^{\text{th}}$  ($\hat{m}^* = 0.56$) percentiles of the estimated lifetime stress index from a DLIM-IM without selection 2 months before through 22 months after last menstrual period (LMP) at weighted stress modifier values in the PROGRESS cohort. The DLIM-IM used 3 degrees of freedom in the modifier and exposure-time dimensions.}
    \label{fig:stress}
\end{figure*}%

We estimated a small critical window of susceptibility and quantified the importance of each stress modifier in contributing to modification of the exposure-time-response function. Figure~\ref{fig:stress} shows the exposure-time-response function at the middle 90\% of the estimated weighted lifetime stress modifier index values. We estimated small critical windows of susceptibility for individuals who experienced moderate to high levels of stress. Specifically, we estimated a window between months 12 to 16 at the $75^{\text{th}}$ percentile of the estimated stress index ($\hat{m}^* = 0.45$) and another window between months 14 to 15 at the $90^{\text{th}}$ percentile of the estimated stress index ($\hat{m}^* = 0.56$). We did not estimate a cumulative effect for any weighted lifetime stress modifier index value (Supplemental Figure 15). We estimated the weight for CRYSIS as 0.283 (SD = 0.096), EPDS as 0.249 (SD = 0.091), STAI as 0.236 (SD = 0.086), and perceived stress as 0.232 (SD = 0.088). Our results suggest that the effect of ambient $\text{PM}_{2.5}$ exposure on HDL 48 months after parturition may be modified by lifetime stress, with each stress modifier contributing similarly to the modification. See Supplemental Figure 16 for a comparison of the estimated exposure-time response curve from the model presented here and models with only one of the modifiers. Further, our analysis demonstrates how the proposed methodology can be used to combine multiple measures of a single construct into a cohesive analysis rather than being forced to arbitrarily combine measures \textit{a priori}, select a single measure, or run repeated analysis with different candidate modifiers. 

\section{Discussion}

In this paper, we propose a Bayesian distributed lag interaction model for multiple continuous modifiers, an extension of the framework for a single continuous modifier in \cite{demateis_penalized_2024}. Our proposed framework simultaneously estimates the exposure-time-response functions varying by a modifier index and the weights of that modifier index. By including a spike-and-slab prior on the weights, we are able to estimate the weights and perform selection on the modifiers, allowing for both identification of the importance of each modifier and selection among candidate modifiers. Our model is versatile and can be applied for both a Gaussian and binomial model. Code for our simulations and analyses can be found as a companion file to this manuscript, and our package is available on GitHub.

Our proposed model fills a methodological gap in the distributed lag framework by allowing for modification of the exposure-time-response function by multiple continuous modifiers. This framework is an extension of the DLIM presented in \cite{demateis_penalized_2024}, which allows for modification by only a single continuous variable. Our proposed framework also builds on index-based models, e.g., weighted quantile sum regression and quantile g-computation, commonly used in environmental health studies \citep{czarnota_assessment_2015, keil_quantile-based_2020}. By creating a weighted modifier index, our proposed model has advantages over models that also allow for modification by multiple modifiers, such as the heterogeneous DLM, because our posterior samples are easily interpretable. Continuous interaction between multiple modifiers and the exposure is a valuable addition to the distributed lag framework.

Through simulation, we considered various modifier index weight scenarios and demonstrated that our proposed DLIM-IM is able to produce more accurate modifier index weight estimates than assuming equal weights. Furthermore, our model is able to estimate the exposure-time-response functions more accurately when compared to the model in \cite{demateis_penalized_2024} with mis-specified fixed modifier index weights. This demonstrates that there is a benefit to using a model that estimates modifier index weights instead of arbitrarily fixing weights \textit{a priori}. Our DLIM-IM with selection produced modifier index weight estimates that were more accurate than those of the DLIM-IM without selection when modifier index weights were truly sparse. 

We applied our proposed model to two data sets to highlight various applications. The first application extended the analysis in \cite{demateis_penalized_2024} by estimating weights of the various Colorado EnviroScreen HSF components to create a data-derived modifier index quantifying neighborhood-level vulnerability and susceptibility. While we estimated similar windows of susceptibility using our DLIM-IM, we were able to further identify which of the HSF components contributed to modification of the exposure-time-response function using PIPs obtained through our modifier selection approach. In the second application, we created a weighted lifetime stress modifier index from four variables all measuring different aspects of stress, including depression during gestation, negative life events, perceived stress during gestation, and anxiety during gestation. Instead of performing multiple analyses with each stress modifier or arbitrarily weighting them together, we were able to estimate a weighted lifetime stress index and identify windows of susceptibility based on the weighted lifetime stress index. 

There are several limitations to our proposed model. First, the weights are constrained to be constant across time. While statistically simple, a model with time-varying weights would be highly parametric and require massive datasets to estimate in practice. The proposed model does allow for modifying variables to change over time. For example, the model could account for a birthing parent who moved during pregnancy. Second, identifiability issues are possible when including highly correlated modifiers, and weight estimation accuracy decreases as the number of potential modifiers increases, as shown in the simulation study. If modifier weight estimation is of interest with a large or highly correlated set of modifiers, we recommend using a pre-selection approach based on existing literature. Third, real data sets often have low signal, and the simulation study demonstrated that modifier index weight estimation may be less accurate. Incorporating prior information on weight can improve estimates.

Our proposed methodology is novel because it allows for modification of the exposure-time-response function by a data-derived index constructed from multiple modifiers. This work extends the distributed lag framework, builds upon index models in environmental health, and fills a critical gap in epidemiology literature.

\section*{Supplemental Material}

We reference additional tables and figures in Sections~\ref{sec.est},~\ref{sec.sim}~\ref{sec.performance},~\ref{sec.simresults},~\ref{sec. analyese},~\ref{sec.COES},~and~\ref{sec.stress}. Code to replicate the simulations and the {\tt R} package {\tt dlimIM} to implement the methods are available at https://github.com/ddemateis/dlimIM.

\section*{Acknowledgments}

This research was supported by National Institutes of Health grants R01ES035735, \\
\noindent R01ES029943, P30ES023515, and R01ES032242.

The authors thank Cantoral Preciado Alejandra de Jes\'us and Martha Mar\'ia T\'ellez Rojo for their field contribution to the PROGRESS study that made this work possible.

These data were supplied by the Center for Health and Environmental Data Vital Statistics Program of the Colorado Department of Public Health and Environment, which specifically disclaims responsibility for any analyses, interpretations, or conclusions it has not provided.

This work utilized the Alpine high performance computing resource at the University of Colorado Boulder. Alpine is jointly funded by the University of Colorado Boulder, the University of Colorado Anschutz, and Colorado State University.

\bibliography{references}

\end{document}